%% file: main.tex
\documentclass[conference]{IEEEtran}
\IEEEoverridecommandlockouts
\usepackage{amsmath,amssymb,amsfonts}
\usepackage{algorithmic}
\usepackage{graphicx}
\usepackage{textcomp}
\usepackage{xcolor}
\usepackage[numbers]{natbib}
\usepackage{booktabs}
\usepackage{multirow,multicol}
\usepackage{bbm}

\def\BibTeX{{\rm B\kern-.05em{\sc i\kern-.025em b}\kern-.08em
    T\kern-.1667em\lower.7ex\hbox{E}\kern-.125emX}}

\input{defs}

\begin{document}

\title{Full Shot Predictions for the DIII-D Tokamak via Deep Recurrent Networks
}


\author{\IEEEauthorblockN{Ian Char}
\IEEEauthorblockA{\textit{Machine Learning Department} \\
\textit{Carnegie Mellon University}\\}
\and
\IEEEauthorblockN{Youngseog Chung}
\IEEEauthorblockA{\textit{Machine Learning Department} \\
\textit{Carnegie Mellon University}\\}
\and
\IEEEauthorblockN{Joseph Abbate}
\IEEEauthorblockA{\textit{Princeton Plasma Physics Laboratory} \\}
\and
\IEEEauthorblockN{Egemen Kolemen}
\IEEEauthorblockA{\textit{Department of Mechanical and Aerospace Engineering} \\
\textit{Princeton Plasma Physics Laboratory} \\
\textit{Princeton University}\\}
\and
\IEEEauthorblockN{Jeff Schneider}
\IEEEauthorblockA{\textit{Machine Learning Department} \\
\textit{Robotics Institute} \\
\textit{Carnegie Mellon University}\\}}

\maketitle

\input{sections/abstract}

\input{sections/introduction}

\input{sections/related}

\input{sections/method}

\input{sections/ablations}

\input{sections/discussion}

\input{sections/acknowledgement.tex}

\bibliography{refs}
\bibliographystyle{IEEEtranN.bst}

\end{document}

%% file: defs.tex
\definecolor{nicegreen}{RGB}{91,226,91}

\newcommand{\indicator}[1]{\mathbbm{1}\{#1\}}

%% file: sections/abstract.tex
\begin{abstract}
Although tokamaks are one of the most promising devices for realizing nuclear fusion as an energy source, there are still key obstacles when it comes to understanding the dynamics of the plasma and controlling it. 
As such, it is crucial that high quality models are developed to assist in overcoming these obstacles.
 In this work, we take an entirely data driven approach to learn such a model. In particular, we use historical data from the DIII-D tokamak to train a deep recurrent network that is able to predict the full time evolution of plasma discharges (or ``shots''). Following this, we investigate how different training and inference procedures affect the quality and calibration of the shot predictions.
\end{abstract}

%% file: sections/introduction.tex
\section{Introduction}

In a wide range of fields, dynamics modeling is a fundamental tool that can be used to gain better understanding of a given system.
Dynamics models are especially useful in the context of control, as they allow for prediction of responses to
system perturbations over time, which can then be used to design and implement control sequences that optimize
for desirable behaviors.

Such benefits are especially apparent in tokamak systems.
Tokamaks are toroidal devices which magnetically confine plasma at high temperatures and pressures
for prolonged periods, during which nuclear fusion reactions occur within the plasma.
The tokamak system is one of the most promising approach to realizing nuclear fusion as an energy source.
While strides are being made to improve the efficiency, stability, and reliability of the system, 
there are crucial control challenges which remain~\citep{humphreys2015novel}.

Since running these devices is extremely expensive, domain experts rely on virtual representations
of the system dynamics, such as simulators and dynamics models. 
Simulators typically rely on first principles and simulate the dynamics via known equations which describe 
the theoretical behavior of the plasma.
However, simulators are prohibitively expensive in terms of time and computation, and despite these costs, they are often still unable to accurately describe the plasma's dynamics.

Concurrently, massive strides have been made in machine learning (ML), where advances in algorithms and modeling architectures
paired with data and compute have allowed for a completely data-driven approach to learning highly accurate models.
This approach is promising for the nuclear fusion setting, and indeed,
numerous recent works have applied ML methods to tokamak modeling \citep{abbate2021data, boyer2021machine, seo2021feedforward, seo2022development, char2023offline}.

In this work, we focus on learning a dynamics model for the DIII-D tokamak, a tokamak in San Diego, California operated by General Atomics. Since the device has been in operation since 1986, we are able to draw from a wealth of previous plasma discharges (or ``shots'') from the device to train a deep recurrent network. A typical shot on DIII-D lasts around 6-8 seconds, with a 1 second ramp up phase, several second flat top phase, and one second ramp down phase. DIII-D also has several real-time and post-shot diagnostics that measure the magnetic equilibrium and plasma parameters with high temporal resolution. We find that learned models are able to predict these measurements for entire shots remarkably well.

We further investigate the impacts of our modelling choices. Along with architecture and training choices, we highlight the importance of uncertainty quantification and explore which methods of forming predictive distributions results in the most calibrated models. As more interest accumulates in control of tokamaks via data-driven models \citep{seo2021feedforward, seo2022development, char2023offline, seo2024avoiding}, we hope that this work provides valuable insights that accelerates prediction and control.

%% file: sections/related.tex
\section{Related Work}

\subsection{Simulators for Tokamaks}

Predictive modeling of the plasma through first principle equations is difficult since different aspects of the plasma evolve at different time scales. 
State of the art simulators solve this problem by evolving these aspects independently \citep{felici2011real}. While these simulators have been useful for exploring different regimes for the plasma \citep{rodriguez2022overview} and making new controls \citep{felici2012non}, they are nevertheless limited in that they require additional external information, such as an estimate for the density at the edge of the plasma. Our learned models are unique from these in that the only information the models require are the settings for the different actuators throughout the shot.

\subsection{System Identification and Machine Learning for Dynamics}

There is a long lineage of methods for inferring behavior of a dynamic system from data.
System identification is one broad categorization of such methods
and can range from ``white'' to ``black'' based on 
how much prior domain knowledge is incorporated.
Whereas white-box models rely strictly on prior knowledge of the relationships
between variables to infer the system parameters, 
black-box methods rely on purely observed data to model their plausible relationships.
We refer the reader to \citet{ljung2010perspectives} and \citet{schoukens2019nonlinear} for a more in-depth, comprehensive survey of existing methods in system identification.

Recently, neural networks (NN) have been widely used for modeling dynamics and have shown substantial success \citep{wang2021physics}. Many of these methods require at least some prior knowledge of the system (i.e. they are gray-box models). 
For example, one may be able to inject prior information into the model using a set of ODEs \citep{mehta2021neural, yin2021augmenting} or PDEs \citep{raissi2018hidden, raissi2019physics}.
There has been recent exciting work applying these classes of models to tokamaks \citep{wang2023hybridizing}; however, in our work, we explore black-box models in which there is no prior knowledge available.

\begin{table*}[ht!]
    \centering
    \begin{tabular}{ccccc}
    \toprule
    \textbf{Group} & \textbf{Representation} & \textbf{Type} & \textbf{Signal} & \textbf{Dimension} \\ \midrule
    \multirow{6}{*}[-2pt]{States} & \multirow{4}{*}[-2pt]{Scalar} & \multirow{2}{*}[-2pt]{Shape} & \multirow{2}{*}[-2pt]{\shortstack{$\kappa$, $a_{\text{minor}}$, Triangularity Top, Triangularity Bottom,\\
    R and Z Coordinates of Magnetic Axis}} & \multirow{2}{*}[-2pt]{6} \\  
    & & & & \\ \cmidrule{3-5}
    & & \multirow{2}{*}[-2pt]{Other} & \multirow{2}{*}[-2pt]{\shortstack{$\beta_N$, Line Averaged Density, Internal Inductance ($li$), $q_0$, $q_{95}$, \\ n1rms, n2rms, n3rms, vloop, wmhd, Differential Rotation~\citep{char2023offline}}} & \multirow{2}{*}[-2pt]{11}\\ 
    & & & & \\ \cmidrule{2-5}
    & \multirow{2}{*}[-2pt]{Profile} & & \multirow{2}{*}[-2pt]{Electron Temperature, Ion Temperature, Density, Rotation, Pressure, $q$} & \multirow{2}{*}[-2pt]{20} \\ 
    & & & & \\ \cmidrule{1-5}
    \multirow{10}{*}[-2pt]{Actuators} & \multirow{10}{*}[-2pt]{Scalar} & \multirow{2}{*}[-2pt]{Beam} & \multirow{2}{*}[-2pt]{Power Injected, Torque Injected} & \multirow{2}{*}[-2pt]{2}\\
    & & & & \\ \cmidrule{3-5}
    & & \multirow{2}{*}[-2pt]{Gas} & \multirow{2}{*}[-2pt]{gasA, gasB, gasC, gasD} & \multirow{2}{*}[-2pt]{4}\\
    & & & & \\ \cmidrule{3-5}
    & & \multirow{2}{*}[-2pt]{Shape} & \multirow{2}{*}[-2pt]{12 Shape Controls} & \multirow{2}{*}[-2pt]{12}\\
    & & & & \\ \cmidrule{3-5}
    & & \multirow{2}{*}[-2pt]{Other} & \multirow{2}{*}[-2pt]{\shortstack{Current Target, Density Target, Toroidal Field}} & \multirow{2}{*}[-2pt]{3}\\
    & & & \\ 
    \midrule
    \multicolumn{5}{c}{\textbf{Total Dimension: 58}} \\
    \bottomrule
    \vspace{2mm}
    \end{tabular}
    \caption{\textbf{List of all state and actuator variables}}
    \label{tab:signals}
\end{table*}
\subsection{Machine Learning in Nuclear Fusion}

There has been a recent surge of interest in applying machine learning methods to predict the state of the plasma within tokamaks. One of the areas with the greatest interest is predicting whether the plasma is in (or is about to be in) an unstable state. \citet{fu2020machine, parsons2017interpretation, rea2019real, boyer2021toward, seo2024avoiding, olofsson2018event, keith2024risk} learn predictive models of whether the plasma will become unstable and use these models to take preventative actions to stabilize the plasma. \citet{char2019offline} also use Bayesian optimization with a similar objective; however, they directly learn the actions to be applied rather than produce a prediction of whether the plasma is unstable or not. 

In terms of learning the evolution of the plasma state, \citet{abbate2021data} learn a deep neural network in order to predict the profiles of the plasma; however, they focus on one step predictions for their model. In contrast, \citet{seo2021feedforward, seo2022development} learn a recurrent neural network to predict scalar states of the plasma. They can use this model to autoregressively predict these states into the future, and they leverage this to plan shots on the KSTAR tokamak. Recently, \citet{char2023offline} used a (non-recurrent) learned model that predicted both scalar and profile states in an autoregressive manner. They used this model as a simulator to train a reinforcement learning agent, which was then deployed on DIII-D. Whereas most of these works focus on the control aspect of dynamics modelling, we do a deeper investigation on the modeling task itself. We hope our exposition on the modeling choices, evaluation techniques, and ablation studies will provide useful insights for future research in dynamics modeling and model-based control for tokamaks.

%% file: sections/method.tex
\section{Method} \label{sec:method}

\subsection{Data}
We begin this section by describing the data used to train our dynamics model. In total, we use 7,884 historical shots from the DIII-D tokamak. We include both the ramp up and flat top phases of each shot, and 
each shot is subdivided into a number of "time steps" 25ms apart from each other.
For each time step, we average the measurements collected 25ms previous to that point in time.

We partition the input signals into two groups: state signals and actuators signals. All of these signals can be found in Table~\ref{tab:signals}. For the state signals, we use 17 different \textit{scalar} states and 6 so-called \textit{profile} states. While scalar states provide a summary statistic of one aspect of the current plasma state with a single scalar, profile states are 1D measurements of the plasma, and in our dataset, they consist of 33 discrete measurements along the minor radius of the cross-section of the tokamak. Following previous works in profile modeling \citep{char2023offline, boyer2021machine}, we choose to lower the dimensionality of the (originally 33-dimensional) profile states via Principle Component Analysis (PCA). In particular, we use four principal components to represent all profiles states except for the pressure and $q$ profile. For these two signals, we use the first two principal components to represent the profile. For actuator signals, we use 21 different scalar values that summarize neutral beam settings, current and density targets, gas settings, and plasma shape control.

We choose to separate these signals into two groups (states and actuators) since we assume that all of the actuators are known a priori (from the perspective of the experiment operator). As such, the model takes as input the current state measurements, the current actuators settings, and the actuator settings 25ms into the future. The model then predicts the change in the state variables for the next 25ms. Once trained, the model can predict many more steps into the future by autoregressively feeding in predicted next states back into the model as inputs.

\begin{figure*}[ht!]
    \centering
    \includegraphics[width=0.85\linewidth]{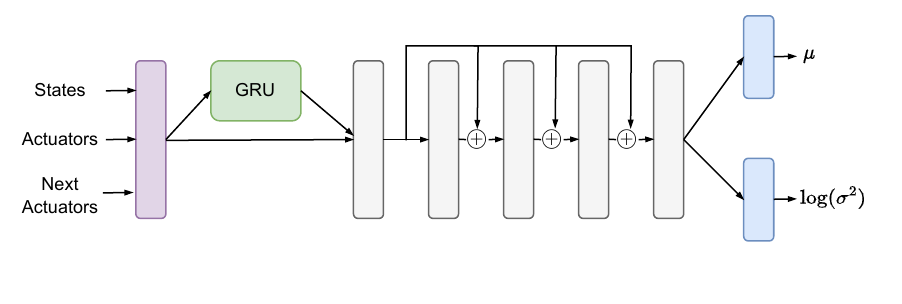}
    \caption{\textbf{Architecture for the Recurrent Model.} The {\color[HTML]{9673A6} encoder} is a single layer MLP which embeds the states, actuators, and next actuators into a 512 dimensional space. This is fed to the {\color[HTML]{82B366} GRU unit} which outputs a 128 dimensional embedding which is concatenated with the original embedding before being fed to the decoder. The {\color[HTML]{6C8EBF} double headed outputs} are single linear layers outputting the mean and log variance of a Gaussian. Note the pluses with circles denote a residual connection.}
    \label{fig:arch}
\end{figure*}

\subsection{Model Architecture and Training}
\label{sec:model}
In designing our model architecture,
we use a recurrent neural network (RNN) with a gated recurrent unit (GRU)~\citep{cho2014properties}. We use 6 hidden layers (including encoder and decoder), each with 512 hidden units and residual connections~\citep{he2016deep}. A visual diagram of the model architecture can be seen in Figure~\ref{fig:arch}. We train our recurrent model with full length shots, the longest of which is 225 time steps. We use a learning rate of $3e-4$ and a weight decay of $0.001$.

For the model output, rather than making point predictions, we have two output heads, where each head predicts the mean and log variance of a Gaussian distribution, respectively. The negative log likelihood (NLL) is computed with this Gaussian prediction, and the model is trained to optimize the NLL loss.
This method of predicting the parameters of a Gaussian distribution via the outputs of a neural network is also known
as a mean-variance network or a probabilistic neural network (PNN)~\citep{nix1994estimating, lakshminarayanan2017simple}, and is one of the most widely used methods of modeling predictive uncertainty.
We extend our discussion on modeling uncertainty in Section~\ref{sec:uq}.

Following \citet{chua2018deep}, we found it essential to place a soft bound on the log variance using a learned lower and upper bound to ensure stability during training. The difference between the upper and lower bound is then added as an additional penalty term to the loss function, encouraging the width of the bounds to be as small as possible.

With the full dataset of shots available, we dedicate 90\% of shots for training, 5\% for validation, and 5\% for testing. The shots are sorted chronologically before the splits are made, and we ensure that the testing shots consist of the 5\% most recent shots. This is essential for testing since experiments (shots) run on the same day tend to be similar. Further, the tokamak is upgraded over time, which alter the dynamics of the plasma.
Hence, enforcing the chronological order not only allows us to test the generalization of the learned dynamics model, but also reflects the realistic test setting that a practitioner will be faced with.

\begin{figure*}[ht!]
    \centering
    \includegraphics[width=\linewidth]{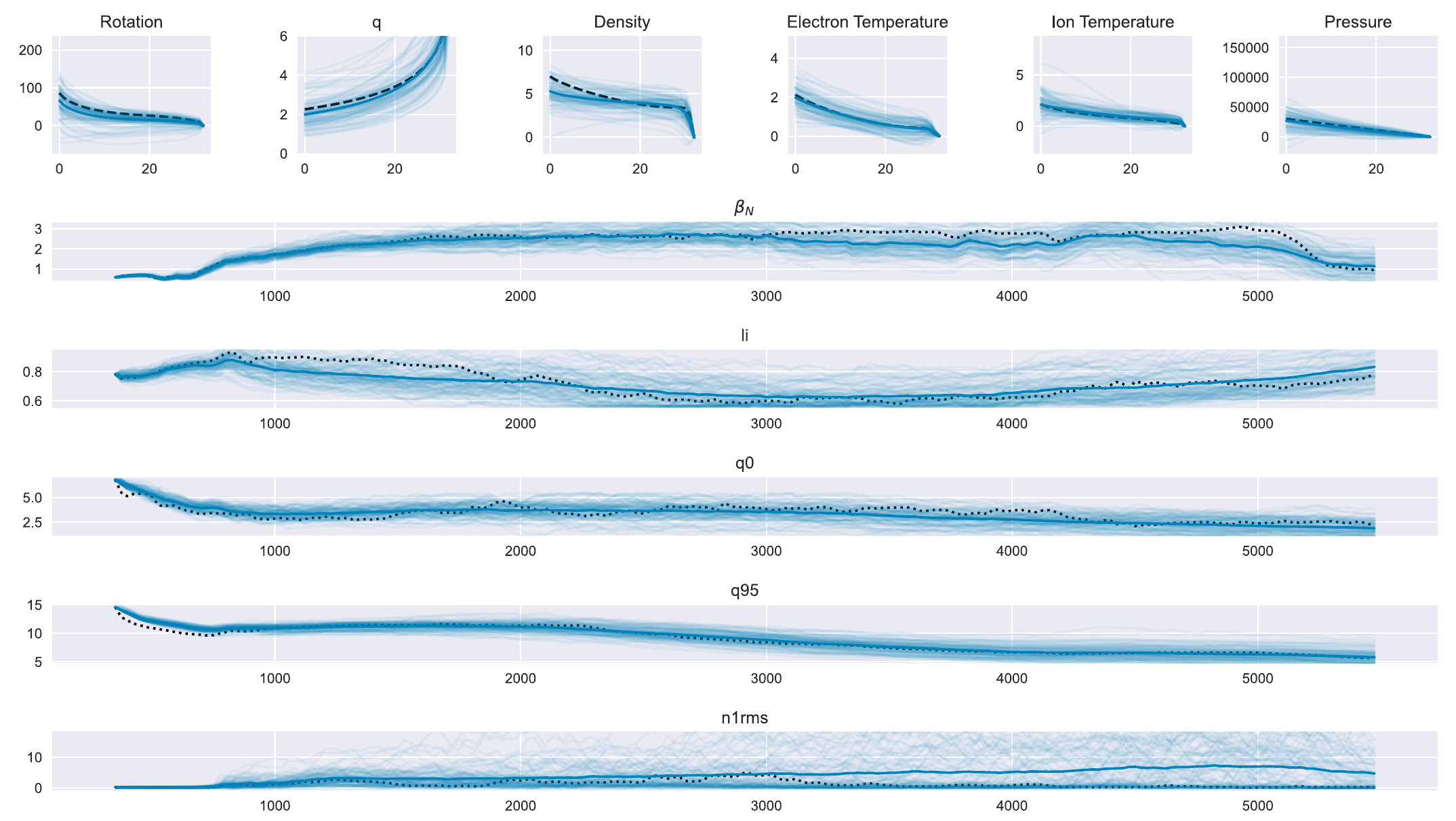}
    \caption{\textbf{Replay of a Test Set Shot} The replay was generated with an ensemble of models which sample from their respective Gaussian distribution at each step. While the model has access to the true actuators throughout the entire shot, it only takes in the first true state and autoregressively predicts the rest. The faded blue lines show one sampled trajectory, while the darker blue line shows the average over the trajectories. The black lines show the true values for the experiment. The top row shows the reconstructed profiles at the last time step. Here, the x-axis is over the minor radius of the tokamak, where 0 is the closest to the magnetic axis and 33 is closest to the wall. The other plots show the scalar values over time. The x-axis shows the time into the shot in ms.}
    \label{fig:replay}
\end{figure*}

\begin{figure*}[ht!]
    \centering
    \includegraphics[width=\linewidth]{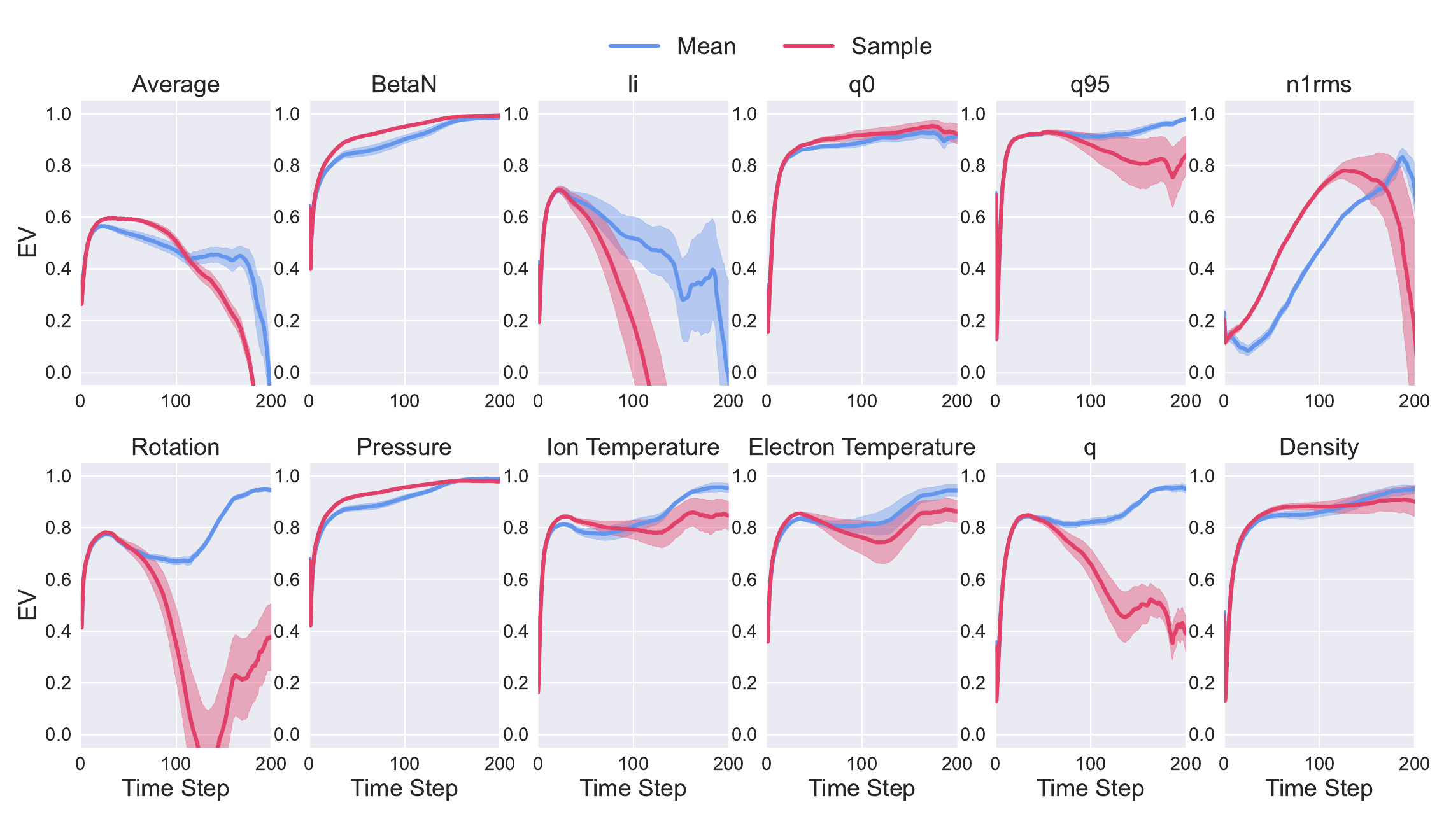}
    \caption{\textbf{Explained Variance per Time Step.} Each of the colored lines show a different way of generating trajectories with the same models. The blue lines simply take the mean of the Gaussian distribution while the red line samples from the Gaussian distribution at every step. Each curve shows the mean over four different models with different random seeds. The shaded area shows the standard error. In the bottom row, we show the EV for the first principle component of the corresponding profiles.}
    \label{fig:eval_ev}
\end{figure*}

\subsection{Evaluation} 

To start, we visualize a ``replay'' of a shot from the test set. That is, we predict the full shot using only the initial state of the plasma and the sequence of actuators. We show the results of this in Figure~\ref{fig:replay}, where we find that the model is able to predict the trend across time for the majority of the plasma's states remarkably well.

To quantitatively evaluate the model's accuracy, we use \textit{Explained Variance} (EV) as an interpretable metric.
This metric is used for the 1D regression setting where, given ground-truth label $y \in \mathbb{R}$ and prediction $\hat{y} \in \mathbb{R}$, the metric is defined as
\begin{align}\label{eq:ev}
    \textrm{EV} := 1 - \frac{\textrm{Var}\left( y - \hat{y} \right)}{\textrm{Var} \left( y \right)}
\end{align}
Here, $\textrm{Var}\left( y - \hat{y} \right)$ and $\textrm{Var} \left( y \right)$ are the empirical variance of the residuals and labels, respectively. Intuitively, this metric shows how much of the variability in the dataset the model can explain, with the maximum (best) score being 1. Since the plasma's state is multi-dimensional, we compute EV for each of the dimensions and for each time step into the future. 

We choose to compute the EV for the difference in the plasma's state at some time step $t$ and the plasma's initial state. With respect to Equation~\ref{eq:ev}, we set $y = s_t - s_0$ and $\hat{y} = \hat{s}_t - s_0$, where $s_t$ is a single dimension of interest in the plasma state at time step $t$, $\hat{s}_t$ is the model's prediction at time step $t$, and $s_0$ is the initial state. We believe that this choice in label better aligns with the task of predicting the evolution of the plasma. Indeed, we found if we measure EV of the plasma state (i.e. we do not subtract $s_0$), EV is nearly perfect for small $t$ since the plasma does not evolve drastically from time step to time step.

For each of the test shots, we evaluate the model by starting prediction at different time steps during the shot (but always seeing the history up to that point), and then autoregressively predicting the remainder of the shot. We assemble the dataset to compute EV using every possible starting time step in all testing shots.
We also compare the EV when sampling from the learned distribution versus using only the mean of the Gaussian. For this sampling method, we sample 30 different trajectories and take the mean over them before computing the EV.

Figure~\ref{fig:eval_ev} displays the EV over time for a select set of scalar quantities, the first component of the profiles, and averaged across all input dimensions. We chose these scalar signals because of their significance. In particular, $\beta_N$ is the normalized ratio between plasma and magnetic pressure and can be used as an indicator of economic performance; $q_0$ and $q_{95}$ are two points along the $q$ (or safety factor) profile, which is an important indicator of stability of the plasma; and n1rms is the root mean squared of magnetic fluctuations for toroidal mode number $n=1$, which can signify an event such as a tearing mode in the plasma. We note that we expect n1rms in particular to be hard to predict.

For each of these plots, EV starts low and grows over time, which can be explained by a number of factors. 
First, there is noise in the system which makes it difficult to predict on the 25ms time scale. However, we hypothesize that the signal starts dominating over the noise after a handful of time steps.
In addition, as the number of time steps starts to grow the variance of $s_t - s_0$ continues to grow (i.e. the denominator of Equation~\ref{eq:ev}), making it easier to achieve higher EV.
In terms of using the mean of the Gaussian versus sampling from it, it seems sampling helps with short term predictions; however, for long term predictions, it appears that using the mean can be more reliable.

%% file: sections/ablations.tex
\section{Ablations}

In this section, we examine a number of choices that we have made for modeling and inference procedures. Our objective is to provide valuable insights that can assist other practitioners in the field when developing dynamic models of plasma evolution in tokamak devices.

\begin{figure*}[ht!]
    \centering
    \includegraphics[width=0.85\linewidth]{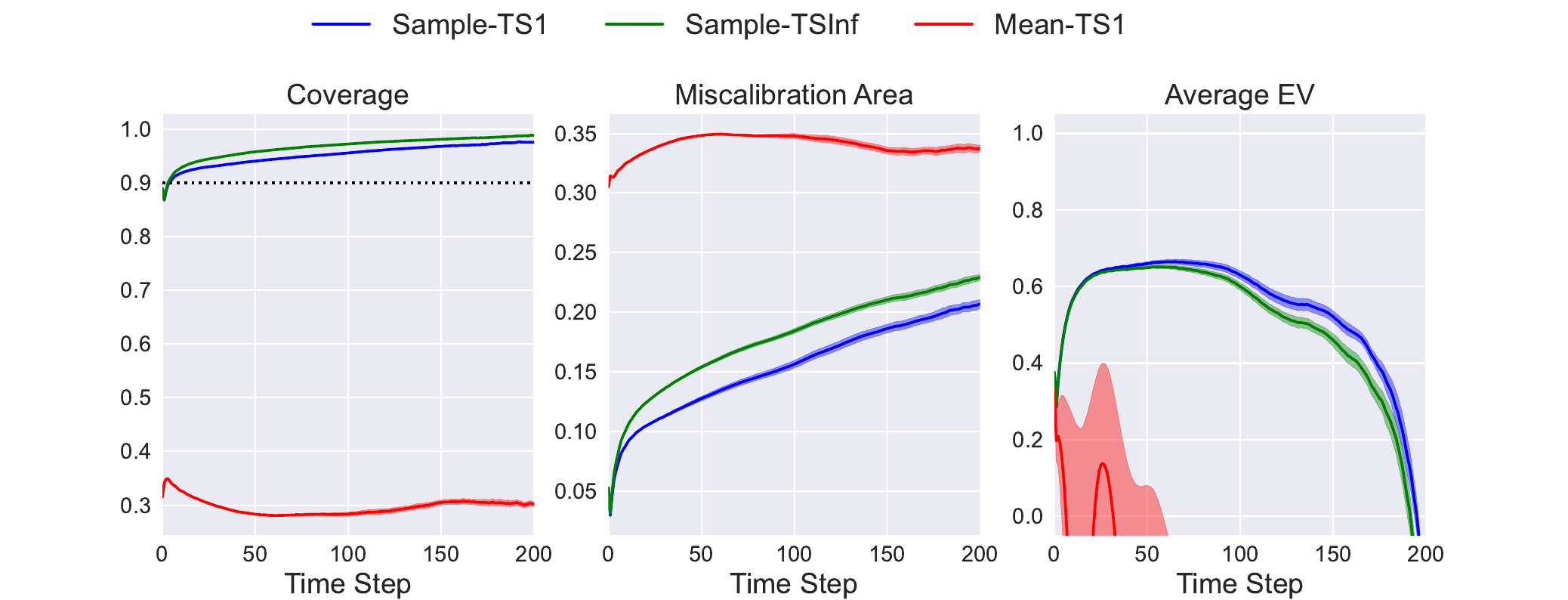}
    \caption{\textbf{Uncertainty Metrics over Time.} The leftmost plot shows coverage of the 90\% prediction interval. Models with good predictive uncertainties should therefore match this 90\% (shown as dotted black line). For the miscalibration area plot, the lower the score, the better calibrated the model is. Each of the metrics is averaged over all output dimensions. Moreover the curves show the mean over four models, and the shaded region shows the standard error.}
    \label{fig:uq}
\end{figure*}

\subsection{Uncertainty Quantification} \label{sec:uq}
Uncertainty quantification is a crucial aspect in any modeling or prediction task, especially in the face of system stochasticity or insufficient data. Adequately modeling uncertainty has been shown to be especially critical for dynamics models when they are leveraged for control~\citep{chua2018deep}.

In our modeling efforts, we account for uncertainty by producing predictive distributions instead of point predictions, and we do so by relying on two different methods: by predicting a Gaussian distribution and by ensembling predictions. These two methods were utilized by~\citet{chua2018deep} to capture the aleatoric uncertainty (the uncertainty inherent in the system) and the epistemic uncertainty (the uncertainty stemming from insufficient data), respectively. The ensemble consists of 4 models, each of which have the same model architecture as described in Section~\ref{sec:model}, but each model was randomly initialized and trained independently~\citep{lakshminarayanan2017simple} on the same dataset.

Inspired by \citet{chua2018deep}, we test three methods of generating predictive distributions from our ensemble of networks, each of which predict Gaussian distributions.
In the first method, which we denote as ``Mean-TS1'', we take the mean prediction of the Gaussian distribution, but sample a new model from the ensemble to generate the next state every step. In the other two methods, we sample from the Gaussian distribution, and we either choose to sample a model from the ensemble every step or every shot. We denote these two methods as ``Sample-TS1'' and ``Sample-TSInf'', respectively.
Because of ensembling and the auto-regressive nature of the model, we do not have a closed form predicted distribution for the plasma's state. Instead, we approximate this distribution with independent Gaussians for each dimension of the plasma's state. The mean and standard deviations are estimated from 30 samples from the dynamics model.

To evaluate the predictive uncertainties, we measure the \textit{Coverage} of a $90\%$ prediction interval (PI) and the so-called \textit{Miscalibration Area}.
At a high level, Coverage is the empirical frequency of observations that fall within a constructed PI. Ideally, if the PI is constructed to capture $1 - \alpha$ probability mass, $(1 - \alpha) \times 100\%$ of the data should fall within this interval.
Concretely, given data points $y_n \in \mathbb{R}$ and $(1 - \alpha)$ intervals $PI_{n, (1 - \alpha)}$ for $n = 1, \ldots, N$, the $(1 - \alpha)$ coverage is defined by
\begin{align*}
    \text{Coverage}_{(1-\alpha)}=\frac{1}{N}\sum_{n=1}^N {\indicator{y_n \in \text{PI}_{n, (1-\alpha)}}}.
\end{align*}
Building on this, the deviation from the expected probability is regarded as miscalibration, and \textit{Miscalibration Area} takes the average deviation over a set of expected probabilities. Given a set of $M$ expected probabilities drawn uniformly from $[0, 1]$: $p_i \sim U[0, 1], i\in[M]$, and the observed $\text{Coverage}_{p_i}$, Miscalibration Area is calculated as 
\begin{align*}
    \frac{1}{M} \sum_{i \in [M]}\mid p_i - \text{Coverage}_{p_i}\mid.
\end{align*}
Much like EV, these metrics are for single dimensional spaces. As such, we compute these metrics for each of the dimensions of the state space and average the results together to produce a single metric. We also compute both metrics for each time step into the future.

Figure~\ref{fig:uq} shows the Coverage, Miscalibration Area (computed with the ``Uncertainty Toolbox'' \citep{chung2021uncertainty}), and EV for the three methods of generating predictive distributions. We see that in the ensemble regime, purely taking the mean of the distribution is detrimental. Not only does the EV suffer, but we also observe extreme overconfidence as shown by the low Coverage (``Mean-TS1'' method in Figure~\ref{fig:uq}).

Looking at the other two methods, we see that both methods are very well-calibrated at the beginning in their short-term predictions. Moreover, there is an improvement in average EV over the single model case (displayed in the top left plot of Figure~\ref{fig:eval_ev}), 
indicating the significance of utilizing an ensemble approach beyond simply quantifying uncertainty. We find that in all aspects, Sample-TS1 dominates over every other method. This aligns with the suggestion given by \citet{chua2018deep}.

As the prediction horizon increases, Miscalibration Area steadily increases, and as evidenced by the Coverage plot, the predictive distributions become increasingly under-confident (i.e. higher than expected coverage). In many cases, one can apply recalibration \citep{kuleshov2018accurate} methods to adjust the calibration; however, we are unaware of any such method for the autoregressive setting.

\begin{table}[]
    \centering
    \begin{tabular}{cc} \toprule
       \textbf{Modelling Choice}  & \textbf{Scaled One-Step MSE} \\ \midrule
        MLP + Gaussian & 1.20 \\ 
        LSTM + Gaussian & 1.05 \\ 
        GRU + Point Prediction & 1.06 \\ 
        \textbf{GRU + Gaussian} & \textbf{1.0} \\ \bottomrule \vspace{2mm}
    \end{tabular}
    \caption{\textbf{One-Step MSE for Model Choices} We scale the MSE by dividing each score by the best MSE that appearing in the table. That is, 1.0 is the best score, while a score of 1.20 means that the MSE achieved was 20\% worse than the best MSE. Each score is the mean over four different seeds.}
    \label{tab:abl}
\end{table}

\begin{figure}[ht!]
    \centering
    \includegraphics[width=1\columnwidth]{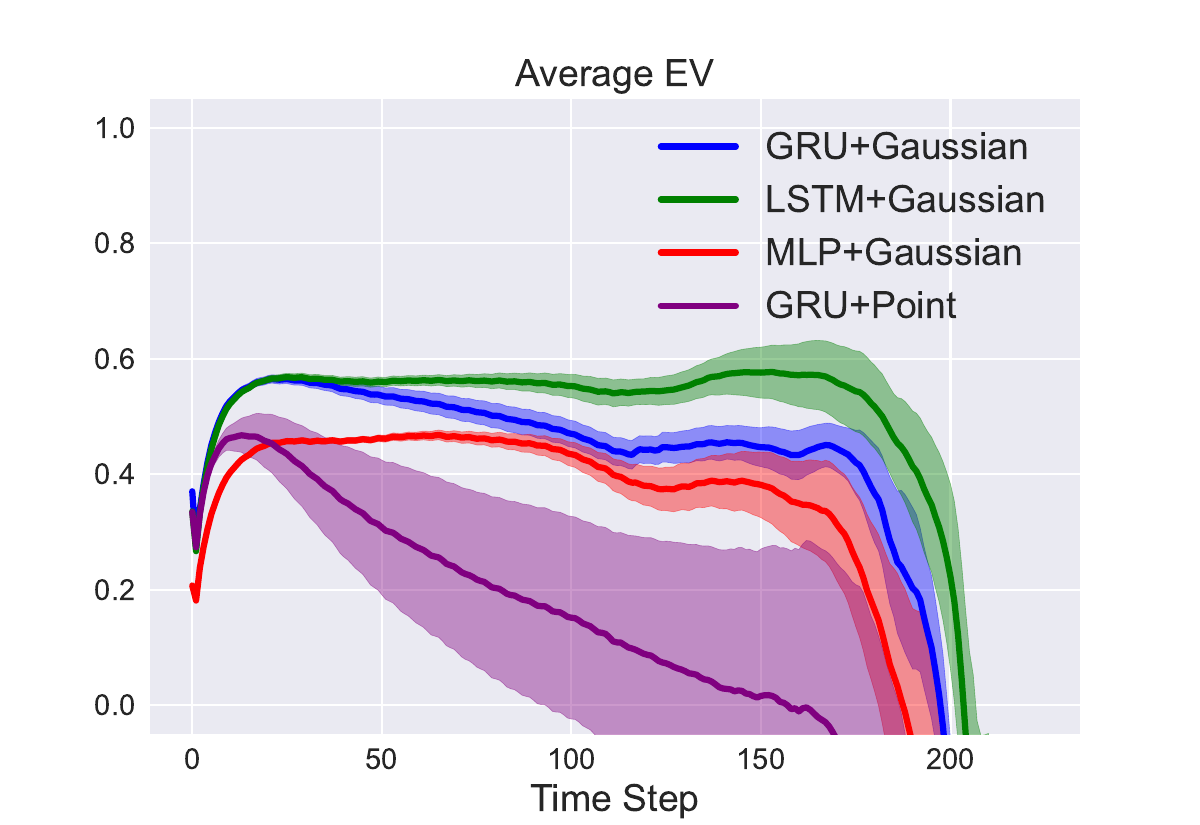}
    \caption{\textbf{Explained Variance Averaged over All Output Dimensions.} Each curve was generated by taking the average over four different trained models. The shaded area shows the standard error. All curves were generated by taking the mean output of the predicted Gaussian distributions (where applicable).}
    \label{fig:abl}
\end{figure}

\subsection{Recurrent Unit}
Next, we consider the impact of the recurrent unit chosen. We consider two alternatives besides the GRU component discussed in Section~\ref{sec:method}: a model with no recurrent unit at all (e.g. an MLP) and a model with an LSTM \citep{hochreiter1997long} unit. From Table~\ref{tab:abl}, one can see that the GRU is superior in terms of single step MSE. Indeed, overall we see that GRU seems to be a good choice when looking at shorter horizon predictions. However, from Figure~\ref{fig:abl} it appears that LSTMs are better when considering a longer time horizon. Therefore, one may want to decide on the best recurrent unit based on the downstream application of the dynamics model. In either case, we see that recurrent units are essential since a standard MLP struggles both with one-step MSE and EV.

\subsection{Point vs Distributional Estimate} 
Lastly, we look at the effect of having the model learn a distribution, as proposed in Section~\ref{sec:method}, as opposed to having the model output a point prediction and training with MSE loss. Interestingly, even though the models that output a point prediction are trained on MSE, they achieve worse MSE on the test set according to Table~\ref{tab:abl}. We hypothesize this is because learning a Gaussian distribution prevents the network from overfitting on the training data. Indeed, we observe that on the training set, models with point predictions achieve roughly 20\% lower MSE when compared with those that output Gaussian distributions. On top of this, Figure~\ref{fig:abl} shows that while models with point predictions have decent EV at first, as one predicts further into the future they achieve worse performance than even models with no recurrent units. 

%% file: sections/discussion.tex
\section{Discussion} In this work, we show that deep recurrent models are a powerful tool that can be used for full shot predictions in tokamak devices. We emphasize that these models were simply given the initial state and the actuators to be applied during the duration of the shot. We encourage the fusion community to leverage data driven models when designing controllers and exploring actuator choices, and we hope that insights shown in this work will prove useful in those pursuits.  

%% file: sections/acknowledgement.tex
\section*{Acknowledgement}

This material is based upon work supported by the U.S. Department of Energy, Office of Science, Office of Fusion Energy Sciences, using the DIII-D National Fusion Facility, a DOE Office of Science user facility, under Awards DE-AC02-09CH1146 and DE-FC02-04ER54698. This work was also supported by DE-SC0021414 and DE- SC0021275 (Machine Learning for Real-time Fusion Plasma Behavior Prediction and Manipulation). Additionally, this work is supported by the National Science Foundation Graduate Research Fellowship Program under Grant No. DGE1745016 and DGE2140739. Any opinions, findings, and conclusions or recommendations expressed in this material are those of the author(s) and do not necessarily reflect the views of the National Science Foundation. Lastly, Youngseog Chung is supported by the Kwanjeong Educational Foundation.

\textbf{Disclaimer} This report was prepared as an account of work sponsored by an agency of the
United States Government. Neither the United States Government nor any
agency thereof, nor any of their employees, makes any warranty, express or
implied, or assumes any legal liability or responsibility for the accuracy,
completeness, or usefulness of any information, apparatus, product, or process
disclosed, or represents that its use would not infringe privately owned rights.
Reference herein to any specific commercial product, process, or service by trade
name, trademark, manufacturer, or otherwise does not necessarily constitute or
imply its endorsement, recommendation, or favoring by the United States
Government or any agency thereof. The views and opinions of authors expressed
herein do not necessarily state or reflect those of the United States Government
or any agency thereof.